\documentclass[aps,prl,floats,showpacs,twocolumn]{revtex4}
\usepackage{graphicx,soul,color}
\sethlcolor{yellow}

\begin{document}
\title{Probing the Superfluid - Mott Insulating Shell Structure\\of Cold Atoms
by Parametric Excitations}
\author{Lilach Goren$^{1}_{}$, Eros Mariani$^{1,2}_{}$, and Ady Stern$^{1}_{}$
  \vspace{1mm}}
\affiliation{$^{1}_{}$ Department of Condensed Matter Physics, The Weizmann Institute of Science, 76100 Rehovot (Israel)\\
$^{2}_{}$ Fachbereich Physik, Freie Universitaet Berlin, Arnimallee 14, 14195 Berlin (Germany)
\vspace{3mm}}
\date{\today}
\begin{abstract}
We study the effect of parametric excitations on systems of
confined ultracold Bose atoms in periodically modulated optical
lattices. In the regime where Mott insulating and Superfluid
domains coexist, we show that the dependence of the energy
absorbed by the system on the frequency of the modulation serves
as an experimental probe for the existence and properties of Mott
insulating-Superfluid domains.
\end{abstract}
 \maketitle

The tunability of ultracold atom gases in optical lattices makes
them ideal candidates for the investigation of quantum
many-particle phenomena in different regimes. As a paradigmatic
example, within the Bose-Hubbard (BH) model, bosonic atoms in
optical lattices have been predicted to undergo a quantum phase
transition from a compressible, phase-coherent superfluid (SF)
phase to an incompressible incoherent Mott insulator (MI) induced
by increasing the ratio of the on-site repulsion $U$ to the
hopping kinetic energy $J$ \cite{Fisher89}. For an infinite
uniform system on a lattice the phase diagram in the
($J/U$,$\mu/U$) plane is composed of lobe-like MI regions with
integer site occupation in an otherwise SF phase
\cite{Fisher89,Freericks94}. By controlling the ratio $J/U$ via
the intensity of the optical lattice, the MI-SF transition has
been experimentally observed \cite{Orzel01,Greiner02} together
with an energy gap of the expected magnitude in the MI phase.
Experiments are always performed in finite inhomogeneous systems
with a fixed total number of atoms subject to an external
confinement which has been numerically established to induce the
coexistence of MI and SF domains
\cite{Jaksch98,Wessel04,Kashurnikov02,Batrouni02}.

The formation of MI shells of different site occupation have been
recently detected \cite{Bloch06} but still, an experimental probe
of the co-existence of SF and MI domains and of their properties
is presently missing. It has been proposed \cite{Eros} that the
presence of MI-SF domains modifies the low-energy excitation
spectrum of the system, finally affecting its thermodynamic
properties.

In this paper we analyze the energy absorption due to a periodic
time modulation of the lattice intensity exploiting the parametric
resonances of the modes of the system. We find this response to be
a probe of the MI-SF domains, particularly at frequencies that
correspond to energies lower than the MI gap. For these
frequencies the energy absorption is only due to excitations of
the SF regions, typically ring-like domains, and is strongly
peaked at frequencies related to the transverse size of the SF
rings. An experimental observation of these peaks may then provide
evidence for the co-existence of MI-SF domains and give an
estimate for their width.

A parametric excitation is a process in which the population of
certain modes is exponentially enhanced as a result of a periodic
temporal modulation of a parameter in the Hamiltonian. Parametric
excitations have been studied theoretically in harmonically
trapped BEC \cite{Castin97} and for condensates in optical
lattices in different spatial dimensions \cite{Ripoll99}. Very
recently a periodic variation of the optical lattice potential
depth was used to probe the response of a 1D condensate across the
MI-SF quantum phase transition \cite{Stoferle04,Esslinger04}.
Here, the energy absorption was measured versus the optical
lattice modulation frequency $\Omega$, showing a broad spectrum in
the SF phase and peaked features in the gapped MI one. Theoretical
works that followed analyzed the parametric instabilities of
Bogoliubov modes \cite{Tozzo1,Tozzo2}. In parallel, very recent
works considered the linear response energy absorption of 1D
systems within the bosonization picture \cite{Giamarchi05} or with
numerical diagonalization of realistic systems
\cite{Kollath06,Jaksch06,Pupillo06}. In the presence of
confinement, the low-energy form of the dynamic structure factor
has been argued to yield signatures of the presence of SF-MI
domains \cite{Pupillo06}.

The BH Hamiltonian describing a system of bosons in a periodic
optical lattice is
\begin{eqnarray}\label{Bose Hubbard H}
  H &=& -J_0\sum_{\langle ij\rangle}\left(
    b_i^\dagger b_j^{} + h.c.\right)+\frac{U_0}{2}\sum_i b_i^\dagger b_i^{} \left(b_i^\dagger
    b_i^{}-1 \right) \nonumber\\
    &+& \sum_i (-\mu+ V_i^{}) b_i^\dagger b_i^{}
\end{eqnarray}
with $b_i^{}$ the annihilation operator of a boson at site $i$. We
assume nearest-neighbour hopping of strength $J_0$, and an on-site
repulsion $U_0$. The values of $J_0$ and $U_0$ are
given by $J_0 =
4/\sqrt{\pi}(V_0/E_R)^{3/4}\text{exp}[-4\sqrt{V_0/E_R}] E_R$ and
$U_0 = \pi^2 a_0/3l (V_0/E_R)^{3/4}E_R$ \cite{Demler2002}, where
$V_0$ and $l$ are the optical lattice depth and spacing, $E_R^{}$ is the
recoil energy and $a_0$ the scattering length.
The chemical potential $\mu$
fixes the particle number and $V_i$ is the potential bias at site
$i$ due to the external confinement.

We consider a two dimensional (2D) atom cloud subjected to an
optical lattice. In the 2D plane the confining potential $V_i$
grows from the center of the cloud towards its boundaries. For
$J_0/U_0$ larger than the critical value inducing the MI-SF
transition, the entire 2D system is a SF while below this critical
ratio concentric alternated 2D ring-like MI and SF domains are
formed \cite{Jaksch98}.

In a compressible SF ring the average occupation per site
$\bar{n}$ is in between the integer occupations of the neighboring
MI domains. Assuming the length of the ring is much larger than
its width $L$, we can view it as a SF stripe closed with periodic
boundaries and consider momentum eigenstates along the stripe. For
weak depletion, in momentum space the BH Hamiltonian (\ref{Bose
Hubbard H}) in the SF stripe reduces, at mean field
\cite{Bogoliubov}, to
\begin{equation}\label{H after MF}
    H = \sum_{\mathbf{k}\neq 0} (\epsilon_\mathbf{k}^0 + \bar{n} U_0 ) a_\mathbf{k}^\dagger a_\mathbf{k}^{} +
\frac{ \bar{n}
U_0}{2}\sum_{\mathbf{k}\neq0}\left(a_{\mathbf{k}}^\dagger
a_{-\mathbf{k}}^\dagger +a_{\mathbf{k}}^{}
a_{-\mathbf{k}}^{}\right)
\end{equation}
where the ground state energy was subtracted and $a_\mathbf{k}^{}
= \sum_j b_j e^{i\mathbf{k}\mathbf{r}_j^{}}$.
Within the tight-binding approximation (valid for the deep lattices implied in the MI domain formation) we have, at small $k$, $\epsilon_\mathbf{k}^0 = J_0 l^2|\mathbf{k}|^2$.\\
The Hamiltonian (\ref{H after MF}) is diagonalized via the
Bogoliubov transformation $c_\mathbf{k}^{} = u_\mathbf{k}^{}
a_\mathbf{k}^{} + v_\mathbf{k}^{} a_{-\mathbf{k}}^\dagger$ with
$u_\mathbf{k}^{2} = 1+v_{\mathbf{k}}^{2}=\frac{1}{2}
\left[\left(\epsilon_\mathbf{k}^{0}+\bar{n}
U_0\right)/E_\mathbf{k}^{0}+1\right]$ and results in
\begin{equation}\label{H diag}
H_0^{} = \sum_{\mathbf{k}\neq0} E_\mathbf{k}^{0}\, c_\mathbf{k}^{\dagger} c_\mathbf{k}^{}\quad ,
\end{equation}
where $c_\mathbf{k}^{\dagger}$ creates a Bogoliubov quasi-particle
of 2D quasi-momentum $\mathbf{k}$ and energy $E_\mathbf{k}^{0} =
\left[(\epsilon_\mathbf{k}^{0})^2 +
2\bar{n}U_0\epsilon_\mathbf{k}^{0}\right]^{1/2}_{}$ .

To the system described by (\ref{H diag}) we introduce a periodic
modulation of the lattice depth $V_0^{}$ of the form $V_0^{}(t) =
V_0^{}\left( 1+A_{}\sin{\Omega t } \right)$, with $A_{}\ll 1$.
Thus, $J_0$ and $U_0$ become time dependent, and to lowest order
in $A_{}$ we have $\epsilon_{\mathbf{k}}^{}(t)
=\epsilon_{\mathbf{k}}^0(1+B_{}\sin{\Omega t }) $ and $U^{}(t)
=U_0(1+C_{}\sin{\Omega t }) $, where $B$ and $C$ are calculated
from the relations below Eq. (\ref{Bose Hubbard H}).
 The resulting Hamiltonian is
\begin{equation}\label{H time dep}
H =\sum_{\mathbf{k} \neq0}\left[ \frac{E_{\mathbf{k}}(t)}{2}\left(c_{\mathbf{k}}^\dagger c_\mathbf{k}^{} + c_{-\mathbf{k}}^\dagger c_{-\mathbf{k}}^{}\right) +\frac{V_{\mathbf{k}}^{}(t)}{2}\left(c_{\mathbf{k}}^\dagger c_{-\mathbf{k}}^\dagger 
+\mathrm{h.c.}\right)\right]
\end{equation}
where $E_{\mathbf{k}}(t) = E_{\mathbf{k}}^0+ \Delta
E_{\mathbf{k}}(t)$, with $\Delta
E_{\mathbf{k}}(t)=\left(\epsilon_{\mathbf{k}}^0/E_{\mathbf{k}}^0\right)[B_{}
\left(\epsilon_{\mathbf{k}}^0+\bar{n}U_0\right)+C_{}\bar{n}U_0]\sin{\Omega
t}$, and
\begin{equation}
V_{\mathbf{k}}^{}(t) =
\frac{\epsilon_{\mathbf{k}}^0}{E_{\mathbf{k}}^0}\bar{n}
U_0(C_{}-B_{})\sin{\Omega t}\quad .
\end{equation}
The Hamiltonian (\ref{H time dep}) is no longer diagonal, and a
time-dependent term creating pairs of excitations with opposite
momenta is introduced, responsible for the parametric resonance.
Following the method of Tozzo {\it et al.}, we introduce the operators $\alpha_{\mathbf{k}}^{} \equiv
c_{\mathbf{k}}^{} \exp\left[i\int dt'
E_{\mathbf{k}}^{}(t')\right]$ and the Heisenberg equations on (\ref{H
time dep}) yield (we set $\hbar =1$)
\begin{eqnarray}\label{original eoms for alpha}
\dot{\alpha}_{\mathbf{k}}^{} = \gamma_{\mathrm{max}}^{} \,\alpha_{-\mathbf{k}}^\dagger \left[e^{-i(\Omega t-2\int{E_{\mathbf{k}}^{}dt})}-e^{i(\Omega t+2\int{E_{\mathbf{k}}^{}dt})}\right]&& \nonumber\\
\dot{\alpha}_{-\mathbf{k}}^\dagger = \gamma_{\mathrm{max}}^{}\, \alpha_{\mathbf{k}} \left[e^{i(\Omega t-2\int{E_{\mathbf{k}}^{}dt})}-e^{-i(\Omega t+2\int{E_{\mathbf{k}}^{}dt})}\right]&&
\end{eqnarray}
where $\gamma_{\mathrm{max}}^{} = (C_{}-B_{})
\epsilon_{\mathbf{k}}^0\bar{n} U_0/2 E_{\mathbf{k}}^0$. To lowest
order in $A$ we may replace $E_{\mathbf{k}}^{}$ by its time
independent part $E_{\mathbf{k}}^0$.

 When the arguments of the
exponents get large, the fast oscillations effectively kill the
time-dependence of $\alpha_{\mathbf{k}}^{}$. In the regime $\Omega
t\gg 1$ this is almost always the case, except on the slowly
oscillating modes with $\left|\Omega -2
E_{\mathbf{k}}^0\right|t\ll 1$ which fulfill the single second
order equation
\begin{equation}\label{eom alpha}
    \ddot{\alpha_{\mathbf{k}}^{}} + i
    \left(\Omega-2E_{\mathbf{k}}^0\right)\dot{\alpha_{\mathbf{k}}^{}}-
    \gamma_{\mathrm{max}}^{2} \alpha_{\mathbf{k}}^{} = 0
\end{equation}
with solution
$\alpha_{\mathbf{k}}^{} = \eta_\mathbf{k}^{+} e^{i\omega_{+}^{} t}+\eta_{\mathbf{k}}^{-} e^{i\omega_{-}^{} t}$, where
\begin{equation}\label{freqs}
    \omega_{\pm} = -\frac{1}{2}\left[(\Omega-2E_{\mathbf{k}}^0)\pm
    \sqrt{(\Omega - 2E_{\mathbf{k}}^0)^2-4\gamma_{\mathrm{max}}^{2}}\right]\; .
\end{equation}
Here $\eta_{\mathbf{k}}^{\pm}$ are set by the initial conditions
$\alpha_{\mathbf{k}}^{}(t=0) = c_{\mathbf{k}}^{}(t=0)$ and
$\dot{\alpha}_{\mathbf{k}}^{}(t=0) = \gamma_{\mathrm{max}}^{}\,
c_{-\mathbf{k}}^\dagger (t=0)$ and result in coherent
superpositions of  $c_{\mathbf{k}}^{}(t=0)$ and
$c_{-\mathbf{k}}^{\dagger}(t=0)$. In a narrow window of energies
$\left|\Omega/2-E_{\mathbf{k}}^0\right|<\left|\gamma_{\mathrm{max}}^{}\right|$
the frequencies $ \omega_{\pm}$ gain an imaginary part and the
occupation of the $\eta_{\mathbf{k}}^{+}$ modes grows
exponentially with time (resonant modes), corresponding to the
resonant excitation of pairs of Bogoliubov modes with opposite
momenta, whose total energy is approximately equal to the external
modulation frequency. Outside this window, the time evolution of
the occupation is periodic. The occupation of
the Bogoliubov modes,
$\langle
c_{\mathbf{k}}^\dagger(t) c_{\mathbf{k}}^{}(t)\rangle$, is then
\begin{eqnarray}\label{finite T occupation}
&&\left(\frac{\gamma_{\mathrm{max}}^{}}{\gamma}\right)^2_{}\sinh^2_{}\left(\gamma t\right)\left[ 2n_{\mathbf{k}}^{}(T)+1\right] +n_{\mathbf{k}}^{}(T) \nonumber\\
&&\left(\frac{\gamma_{\mathrm{max}}^{}}{\gamma}\right)^2_{}\sin^2_{}\left(\gamma t\right)\left[ 2n_{\mathbf{k}}^{}(T)+1\right] +n_{\mathbf{k}}^{}(T)
\end{eqnarray}
in the regimes $4\gamma_{\mathrm{max}}^{2} >
(\Omega-2E_{\mathbf{k}}^0)^2$ and $4\gamma_{\mathrm{max}}^{2} <
(\Omega-2E_{\mathbf{k}}^0)^2$ respectively, where $\gamma \equiv
\left[
|\gamma_{\mathrm{max}}^{2}-\frac{1}{4}(\Omega-2E_{\mathbf{k}}^0)^2_{}|\right]^{1/2}_{}$
and $n_{\mathbf{k}}^{}(T) =[e^{E_{\mathbf{k}}^0/k_{\mathrm{B}}^{}
T}-1]^{-1}$ is the Bose distribution. This result is correct as
long as the occupation of the excited modes (the condensate
depletion) is small such that (\ref{H after MF}) and (\ref{H time
dep}) hold. The exponential growth of the resonant-modes
occupation is compatible with the result of Tozzo et al.
\cite{Tozzo2}.

In order to calculate the energy absorbed by the system due to the
time dependent modulation of the lattice depth, we consider a
modulation that is turned on at $t=0$ and switched off at time
$t=\tau$, bringing back the Hamiltonian of the system to (\ref{H
diag}). The energy absorbed by the system during the time
$0<t<\tau$ is
\begin{equation}\label{energy absorption}
    \mathcal{E}(\tau) = \sum_{\mathbf{k}\neq0} E_{\mathbf{k}}^{} \left[\langle c_{\mathbf{k}}^\dagger(\tau) c_{\mathbf{k}}^{}(\tau)
    \rangle - \langle c_{\mathbf{k}}^\dagger(0) c_{\mathbf{k}}^{}(0)\rangle\right]
\end{equation}
where 
we perform a thermal average with the unperturbed Hamiltonian
$H_0^{}$. The absorption is then obtained by inserting
(\ref{finite T occupation}) into (\ref{energy absorption}), with
the summation over $\mathbf{k}$ limited to the range of energies
where oscillations are slow.

The absorption is strongly affected by the density of states (DOS)
of the system, which is crucially dependent on the shell
structure. Indeed, if the width of the ring $L$ is much smaller
than its length $L_x^{}$, transverse quantization of the 2D
momenta yields a spectrum composed of separate 1D branches,
labelled by the integers $m$. The $m_{}^{th}$ branch has a
dispersion
\begin{equation}\label{bog dispersion branches}
E_{k,m}^0 \simeq c\,\left[k^2+(\frac{\pi m_{}}{L})^2\right]^{1/2}_{}
\end{equation}
where $k$ is the 1D wavenumber along the ring and we considered
the low-energy phononic dispersion with sound velocity
$c=\left(2\bar{n}U_0J_{0}^{}l^{2}_{}\right)^{1/2}_{}$. The DOS is
characterized by Van-Hove singularities near $k=0$ of the form
$\nu(E)\sim E/\left[E^{2}_{}-E_{g,m}^{2}\right]^{1/2}_{}$ where $E_{g,m} =
E_{k=0,m}^0$. In addition, in the presence of the SF rings, 1D
surface modes exist at the interfaces between domains, with a
linear dispersion at low momenta \cite{Eros}. Their DOS is however
smooth and shows no singularity, which makes their detection
difficult with the technique proposed here.

\begin{figure}[h]
\begin{center}
\includegraphics[width=6.5cm,height=4.7cm]{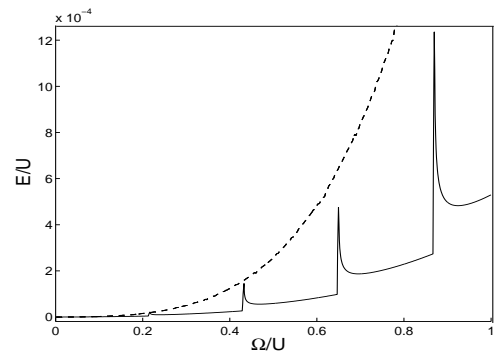}
\end{center}
\caption{The total energy absorption after a modulation of the
lattice intensity $V_0^{}$ with $A_{}= 0.001$ during $10\
\text{msec}$, plotted as function of the modulation frequency
$\Omega/U_0$ for the mixed MI-SF regime at $U_0/J_0=45$ (solid
line) for the inner ring, of average density $\bar{n}=1.5$
particles per site, and for the 2D SF phase at $U_0/J_0=20$
(dashed line) and average density of $\bar{n}=1.5$. We considered
$l$=257 nm and $L=10\, l$.}\label{Fig1}
\end{figure}

We now examine the dependence of the energy absorption
(\ref{energy absorption}) on the modulation frequency $\Omega$.
The typical parameters used in recent experiments
\cite{Stoferle04} are $\tau\sim 30$ ms and $\Omega\leq8$ kHz such
that $\Omega\tau\gg 1$ and the occupation of the modes is
determined by Eq. (\ref{eom alpha}). At a given $\Omega$, modes
with energies in a "resonant window" of width
$2\gamma_{\mathrm{max}}^{}$ around $\Omega/2$ are exponentially
populated in time. Thus, by varying $\Omega$ and considering the
energy absorbed by the system we can get information about its
spectrum. For example, in the presence of SF rings, changing
$\Omega$ in the vicinity of $2E_{g,1}^{}$ the resonant window
moves across the van Hove singularity and the absorption shows a
peak. The absorption is then suppressed as $\Omega$ is further
increased past the singularity. A similar enhancement
is predicted every time the
modulation frequency meets the condition
$2(E_{g,m}^{}-|\gamma_{\mathrm{max}}^{}|)<\Omega<2(E_{g,m}^{}+|\gamma_{\mathrm{max}}^{}|)$
for some $m$. Fig. \ref{Fig1} shows the energy absorbed by the
system versus $\Omega$ for a modulation of amplitude $A_{}\sim
0.001$ in the lattice intensity $V_0^{}$. For a typical $J_0
\approx U_0/45\approx 0.4 \,\text{KHz}$ (characteristic to the
$\bar{n}=2$ MI lobe) this corresponds to the amplitudes
$B_{}=0.01$ in the hopping strength and $C_{}\sim 0.001$ in the
on-site repulsion.

The plot in Fig. \ref{Fig1} is the main result of this paper; a
measurement of the energy absorption as function of the modulation
frequency supplies a direct evidence for the existence of narrow
SF shells, as can be seen from the difference between the dashed
and the solid lines. We point out that the enhanced absorption
shown in the latter is a result of the divergence in the DOS
together with the fact that the exponent
$\gamma_{\mathrm{max}}^{}$ does not vanish at $k=0$. In the
low-energy regime $2\bar{n}U_0\gg\epsilon_{k,m_{}}^0$ we get
$\gamma_{\mathrm{max}}^{} \simeq (C-B) E_{k,m_{}}^0/4$ which is
finite for $k=0$ if $m\neq 0$.

 The energy difference between the
transverse branches resulting from one SF ring, $\Delta
E=E_{g,m_{}}-E_{g,m-1_{}} \simeq \pi c/L$ is inversely
proportional to its width $L$. Therefore the spacing between
adjacent absorption peaks in Fig. \ref{Fig1} is a signature of the
ring-like SF domains, and their typical width may be estimated
from it, given $J_0, U_0$ and $l$.

This result may be compared with the expected energy absorption in
the absence of SF rings. In case $J_0/U_0$ is very small, the
width of the SF rings gets smaller than one lattice spacing, and
the SF part of the system effectively disappears. Then, the only
modes that can be excited are above the MI gap and no signature
appears in low energy absorption. In the opposite regime, when
$J_0/U_0$ is larger than the critical ratio for the MI-SF
transition, the entire 2D system is SF with a smooth DOS. The
dashed line in Fig. \ref{Fig1} shows the energy absorption as
function of the modulation frequency in this regime.

 For a 3D
cloud of atoms the arguments above hold in essentially the same
form. Again, the presence of SF shells with thickness $L$ yield
several 2D Bogoliubov branches with interband separation
proportional to $1/L$. In this case, the 2D nature of the modes
does not lead to van Hove singularities but rather to a staircase
profile in the DOS. As a result, the absorption vs. $\Omega$ will
not show peaks, as in Fig. \ref{Fig1}, but steps, from the
existence and spacing of which one can infer the presence of
domains and their size. Again, in contrast, a pure 3D superfluid
phase would show a smooth absorption spectrum.

We now analyze the consistency of our results and their
applicability to experiments. We choose the parameter values to be
in a regime close to that described in \cite{Jaksch98}. The recoil
energy for Sodium atoms in an optical lattice of wavelength $2l =
514\, \mathrm{nm}$ is $E_R = 2\pi\hbar\times 32 \mathrm{kHz}$. For
the generation of two MI rings we need
$V_0\simeq 15\, E_R^{}$,
yielding $U_0\simeq 0.4\, E_R^{}$ and $J_0\simeq 0.01\, E_R^{}$. The single band BH
approximation holds, since
$\frac{1}{2}U_0\bar{n}(\bar{n}-1)\ll \hbar \nu$, where $\hbar \nu
=\sqrt{4 E_R^{} V_0^{}}$ is the excitation energy to the first
excited Bloch band.
A typical ring of width $L\simeq 10\, l$ and $\bar{n}=1.5$ results
in a transverse gap of $E_{g,1}^{}\simeq 0.1\, U_0$,
allowing to probe about 10 transverse branches below the
MI gap $U_0$. 
The visibility of the absorption
peaks is strongly related to the duration of the parametric
modulation.
An estimate for the minimal duration required for good resolution
of the peaks is $\tau \sim
\frac{1}{2}\hbar\gamma_{\mathrm{max}}^{-1}(E_{g,1})\sim 10
\,\mathrm{msec}$, feasible in current experimental setups.

The enhancement in the
population of modes at zero temperature is triggered by
the presence of the zero point fluctuations, that supply a "quantum seed". At finite temperature these are augmented by thermal fluctuations.
A second important fact enabling this amplification is the
bosonic character of the atoms. If we consider the "fermionic
version" of our hamiltonian (\ref{H after MF}), i.e. the
BCS mean-field one, and proceed with the periodic modulation, no
exponential growth of the occupation is expected due to the
fermionic nature of the excitations.

The sharp peaks in Fig. \ref{Fig1} result from the confinement of
the superfluid to narrow and long strips and from the long-living
excitation modes that characterize superfluids. To separate the
contribution of these two sources, we consider the effect of
viscosity on the parametric resonances in a narrow fluid strip. We
concentrate on the parametric excitation of sound waves
when the sound velocity is modulated through a periodic variation
of the effective mass $M(t) = M_{0}(1+B\sin{\Omega t})$ and the
interaction constant $U(t)=U_0(1+C\sin{\Omega t})$. The equation
of motion for the density fluctuation is
\begin{equation}\label{eom dissipation}
    \ddot{\rho}_k+ \zeta_k(t) \dot{\rho_k}+\omega^2_k(t)\rho_k=0
\end{equation}
where  $\omega^2_k(t)=c_s(t) k$, $c_s(t) \propto
\sqrt{U(t)/M(t)}$, $\zeta(t)\equiv \dot{M}(t)/M(t)+ \eta k^2$, and
$\eta$ is the fluid viscosity. The transformation $\rho_k(t)
=y_k(t)\exp{[-\int_0^t{\zeta_k(t')/2}dt']}$ maps Eq.~(\ref{eom
dissipation}) onto a parametrically excited oscillator
\begin{equation}
  \ddot{y_k}+\tilde{\omega}^2_{k,0} \left[ 1+ \tilde{A}\sin{\Omega (t-t_0)}\right] y_k =0
\end{equation}
 where $\tilde{A}=\left[[C-(1-\Omega^2/2\omega_{k,0}^2)B]^2+(\eta k^2
\Omega B/2\omega_{k,0}^2)^2 \right]^{1/2}$ and
$\tilde{\omega}^2_{k,0}=\omega_{k,0}^2-(\eta k^2/2 )^2$ to lowest
order in $B$ and $C$. The resonant $y_k's$ are exponentially
enhanced, $y_k(t)\sim\exp{[\tilde{\gamma}_{max} t]}$ with
$\tilde{\gamma}_{max} \simeq
\frac{1}{4}\tilde{A}\tilde{\omega}_{k,0} $. The resulting dynamics
of the sound modes has the form $\rho_k\sim
\exp{[(\tilde{\gamma}_{max}-\eta k^2/2 )t]}\exp [{-B \sin{\Omega
t}/2}]$.

Density fluctuations will grow exponentially in time only if the
amplitude of the modulation is large enough such that
$\tilde{\gamma}_{max}$ is real and satisfies
$\tilde{\gamma}_{max}>\eta k^2/2$. Furthermore, the width of the
energy absorption peaks is $2\tilde{\gamma}_{max}$. Thus, when the
amplitude of the modulation is large enough for the formation of
the peaks, the peaks acquire a finite width, which for the $m$'th
peak must be larger than $\eta (\pi m)^2/2L^2$. When the viscosity
is large enough such that the width of the peaks is larger than
their spacings, the peaks are fully smeared.

In conclusion, we considered an ultracold cloud of Bose atoms
subjected to a periodically modulated optical lattice which
induces parametric amplification of its excitations. The energy
absorption as a function of the modulation frequency may serve as
a tool to detect features in the DOS of the system, that are
induced by a transition from a uniform SF phase for weak lattice
intensities to a mixed MI-SF regime for stronger ones. This
enables the experimental detection of the coexistence of MI-SF
domains and the measurement of their typical size. In addition, we
demonstrated the broadening of absorption peaks due to damping in
normal fluids which is absent in case the domains are superfluids.

We acknowledge the support from the Feinberg School of the
Weizmann Institute of Science, the US-Israel BSF and the Minerva
foundation.

 \emph{Note added:} During the final stage
of preparation of this manuscript two experiments appeared testing
the shell structure with local probes \cite{Folling06,Campbell06}.


\begin{thebibliography}{99}


\bibitem{Fisher89}  M. P. A. Fisher et al., Phys. Rev. B \textbf{40}, 546
(1989).

 \bibitem{Freericks94} J. K. Freericks and H. Monien, Europhys. Lett.
\textbf{26}, 545 (1994); D. van Oosten et al., Phys. Rev. A
\textbf{63}, 053601 (2001); T. D. Kuehner and H. Monien, Phys.
Rev. B \textbf{58}, R14741 (1998); N. Elstner and H. Monien, ibid
\textbf{59}, 12184 (1999)

%
%
%

\bibitem{Orzel01}  C. Orzel et al., Science \textbf{291}, 2386 (2001).

\bibitem{Greiner02}  M. Greiner et al., Nature \textbf{415}, 39 (2002).

\bibitem{Jaksch98}  D. Jaksch et al., Phys. Rev. Lett. \textbf{81}, 3108 (1998).

\bibitem{Wessel04} S. Wessel et al., Phys. Rev. A \textbf{70}, 053615 (2004).

\bibitem{Kashurnikov02}  V. Kashurnikov et al., Phys. Rev. A \textbf{66}, 031601(R) (2002).

\bibitem{Batrouni02}  G. G. Batrouni et al., Phys. Rev. Lett. \textbf{89}, 117203 (2002).

\bibitem{Bloch06}  F. Gerbier et al., Phys. Rev. Lett. \textbf{96}, 090401 (2006).

\bibitem{Eros}  E. Mariani and A. Stern, Phys. Rev. Lett. \textbf{95}, 263001 (2005).

%

%
%

\bibitem{Castin97}  Y. Castin and R. Dum, Phys. Rev. Lett. \textbf{79}, 3553 (1997).

\bibitem{Ripoll99}  J. J. Garc\'{i}a-Ripoll et al., Phys. Rev. Lett. \textbf{83}, 1715
(1999); Yu. Kagan and L. A. Maksimov, Phys. Rev. A \textbf{64},
053610 (2001).



\bibitem{Stoferle04}  T. St\"{o}ferle et al., Phys. Rev. Lett. \textbf{92}, 130403 (2004).

\bibitem{Esslinger04} C. Schori et al., Phys. Rev. Lett. \textbf{93}, 240402 (2004).

\bibitem{Tozzo1}  M. Kr\"{a}mer et al., Phys. Rev. A \textbf{71}, 061602(R) (2005).

\bibitem{Tozzo2}  C. Tozzo et al., Phys. Rev. A \textbf{72}, 023613 (2005).

\bibitem{Giamarchi05}  A. Iucci et al., cond-mat/0508054.

\bibitem{Kollath06}  C. Kollath et al., cond-mat/0603721.

\bibitem{Jaksch06}  S. R. Clark and D. Jaksch, cond-mat/0604625 (2005).

\bibitem{Pupillo06} G. Pupillo et al., cond-mat/0602240.

\bibitem{Demler2002}  E. Demler and F. Zhou, Phys. Rev. Lett. \textbf{88}, 163001 (2002).

\bibitem{Bogoliubov}  N. N. Bogoliubov, J. Phys. (USSR) \textbf{11}, 23 (1947)

\bibitem{Folling06} S. Foelling et al., cond-mat/0606592

\bibitem{Campbell06} G. K. Campell et al., cond-mat/0606642

\end{thebibliography}
\end{document}